\documentstyle[12pt]{article}
\advance\textheight by \topskip
\makeatletter

\@addtoreset{equation}{section}
\makeatother

\newcommand{\BE}{\begin{equation}}
\newcommand{\EE}{\end{equation}}
\newcommand{\BA}{\begin{eqnarray}}
\newcommand{\EA}{\end{eqnarray}}

\def\no{\nonumber}
\def\bi{\bibitem}

\def\ap{\alpha ' }
\def\B{\beta}
\def\BH{\beta_{\scriptscriptstyle H}}
\def\th{\vartheta}
\def\T2{|T|^2}
\def\Eb{\overline{E}}
\def\p{\partial}
\def\a{\alpha}
\def\D#1{D#1-$\overline{\textrm{D#1}}$}
\def\Cm{C_{\scriptscriptstyle -1}}
\def\Dm{D_{\scriptscriptstyle -1}}
\def\Ch{C_{\scriptscriptstyle - \frac{1}{2}}}
\def\Dh{D_{\scriptscriptstyle - \frac{1}{2}}}
\def\v{{\cal V}}
\def\vd{{\cal V}_{d}}
\def\vp{{\cal V}_{p}}
\def\A{A}
\def\Ap{A'}
\def\db{\overline{d}}
\def\Bn{\beta_{{\bf n}}}

\input epsf

\begin{document}

\rightline{KUNS-1831}
\rightline{hep-th/0303236}

\vspace{.8cm}
\begin{center}
{\large\bf Finite Temperature Systems of Brane-Antibrane on a Torus
}

\vskip .9 cm

{\bf Kenji Hotta,}
\footnote{E-mail address: khotta@gauge.scphys.kyoto-u.ac.jp}

Department of Physics, Kyoto University, Kyoto 606-8502, JAPAN
\vskip 1.5cm

\end{center}
\vskip .6 cm
\centerline{\bf ABSTRACT}
\vspace{-0.7cm}
\begin{quotation}

In order to study the thermodynamic properties of brane-antibrane systems in the toroidal background, we compute the finite temperature effective potential of tachyon $T$ in this system on the basis of boundary string field theory. We first consider the case that all the radii of the target space torus are about the string scale. If the {\D{p}} pair is extended in all the non-compact directions, the sign of the coefficient of $\T2$ term of the potential changes slightly below the Hagedorn temperature. This means that a phase transition occurs near the Hagedorn temperature. On the other hand, if the {\D{p}} pair is not extended in all the non-compact directions, the coefficient is kept negative, and thus a phase transition does not occur. Secondly, we consider the case that some of the radii of the target space torus are much larger than the string scale and investigate the behavior of the potential for each value of the radii and the total energy. If the {\D{p}} pair is extended in all the non-compact directions, a phase transition occurs for large enough total energy.

\end{quotation}

\normalsize
\newpage

\section{Introduction}
\label{sec:Intro}

Non BPS configurations of branes such as brane-antibrane pairs and non BPS D-branes have received attention recently (for a review see e.g. \cite{Ohmori}). Open strings on these branes have a tachyonic mode, and these configurations are unstable. The tachyon potential on these configurations is calculated by using string field theory. For example, the tachyon potential on the {\D{p}} system in type II string theory was computed on the basis of boundary string field theory (BSFT) \cite{BSFT1} \cite{BSFT2} by Minahan and Zwiebach \cite{tachyon1} and by Kutasov, Marino and Moore \cite{tachyon2}. It is given by
\BE
  V(T) = 2 \tau_p \v \exp (-8 \T2),
\EE
where $T$ is a complex scalar tachyon field, $\v$ is the volume of the system that we are considering, and $\tau_p$ is the tension of a Dp-brane, which is defined by
\BE
  \tau_p = \frac{1}{(2 \pi)^p {\ap}^{\scriptscriptstyle \frac{p+1}{2}} g_s},
\label{eq:tension}
\EE
where $g_s$ is the coupling constant of strings. This potential has the minimum at $|T| = \infty$, and its qualitative features agree with Sen's conjecture \cite{Senconjecture}.

Although the tachyon potential is unstable at $T=0$, there is a possibility that $T=0$ becomes a stable minimum at high temperature. There are some approaches to the statistical mechanics of open strings on the {\D{p}} system. Danielsson, G\"{u}ijosa and Kruczenski evaluated the finite temperature effective potential when there is only the tachyon field \cite{LowTtach}. Huang attempted to evaluate the finite temperature effective potential based on BSFT of bosonic strings \cite{Huang1} and then investigated a finite temperature system of the parallel brane and antibrane with a finite distance \cite{Huang2}. However, in these models, we cannot see the behavior of the system near the Hagedorn temperature, which is the maximum temperature of perturbative strings.

In the previous work \cite{Hotta4}, we have computed the finite temperature effective potential of the brane-antibrane systems near the Hagedorn temperature on the basis of the work of Andreev and Oft \cite{1loopAO}. In the {\D{9}} case, the sign of the coefficient of $\T2$ term of the potential changes slightly below the Hagedorn temperature from negative to positive as the temperature increases, while the coefficient is kept negative in the {\D{p}} case with $p \leq 8$. Therefore, we conclude that not a lower dimensional brane-antibrane pair but a spacetime-filling {\D{9}} pair is created near the Hagedorn temperature. This is the result in the case that the target space is non-compact.

A considerable number of studies have been made on the application of the brane-antibrane system to cosmology \cite{inflation1} \cite{inflation2} (for a review see e.g. \cite{Gibbons}). It is important to investigate the brane-antibrane systems from the cosmological viewpoint because the temperature is extremely high in the early universe, and brane-antibrane pairs might exist in this period. Then the universe expands inflationary, because the tension energy of the brane-antibrane pairs can provide an effective cosmological constant. In the early universe, the size of the universe is extremely small and the topology of the universe is also very important. This is because Strings and branes can wrap the compact directions of target space and they are sensitive to its topology. The purpose of this paper is to investigate the brane-antibrane system at finite temperature in the toroidal background, which is the simplest non-trivial one.

This paper is organized as follows. In \S \ref{sec:free} we evaluate the free energy of open strings on the {\D{p}} system in the toroidal background on the basis of BSFT. We must compute the finite temperature effective potential by using the microcanonical ensemble method, because we cannot trust the canonical ensemble method near the Hagedorn temperature as we will explain later. Then we investigate the behavior of the free energy in the complex $\B$-plane in \S \ref{sec:comT} and compute the finite temperature effective potential in the microcanonical ensemble method in \S \ref{sec:poten}. \S \ref{sec:conclusion} presents our conclusions and discussions.

\section{Free Energy of Open Strings}
\label{sec:free}

It is well-known that we must compute the statistical variables of strings near the Hagedorn temperature by using the microcanonical ensemble method \cite{Efura}. This is because we cannot trust the canonical ensemble method near the Hagedorn temperature from the following reason. The partition function $Z(\B)$ is given by the Laplace transformation of the density of states $\Omega (E)$:
\BE
  Z(\B) = \int_{0}^{\infty} dE \ \Omega (E) e^{- \B E}.
\label{eq:Lap}
\EE
We can expect that the canonical ensemble method gives the same quantities with that of the microcanonical ensemble method if the integrand has a sharp peak. However, the density of states of strings behave as
\BE
  \Omega (E) \sim e^{\BH E},
\EE
for large $E$, where $\BH$ is the inverse of the Hagedorn temperature, and the integrand has no sharp peak near the Hagedorn temperature. Thus, we must compute the finite temperature effective potential by using the microcanonical ensemble method. $\Omega (E)$ can be obtained from the inverse Laplace transformation of $Z(\B)$:
\BE
  \Omega (E) = \int_{L-i \infty}^{L+i \infty}
    \frac{d \B}{2 \pi i} Z(\B) e^{\B E}.
\label{eq:inLap}
\EE
The density of states has been derived from this formula in the case of closed strings \cite{Tan2} \cite{Hotta2} and in the case of open strings on D-branes \cite{Thermo}, and in the case of open strings on the {\D{p}} system \cite{Hotta4}. We will apply this method in the case of open strings on the {\D{p}} system in the toroidal background. Since $Z(\B)$ is related to the free energy $F( \B )$ as
\BE
  Z(\B) = \exp [ - \B F( \B ) ],
\label{eq:partitionfree}
\EE
we will begin by evaluating the free energy of open strings.

Let us make a comment here. One may wonder why we use such an indirect method to compute the density of states. The density of states is defined by
\BE
  \Omega (E) = tr \delta \left( E - {\cal H} \right),
\EE
where $E$ is the total energy and ${\cal H}$ is the Hamiltonian. It is difficult to take the trace over the states of strings directly. If we use complex integral representation of the delta function and the definition of $Z( \B )$, which is given by
\BE
  Z( \B ) = tr e^{- \B {\cal H}},
\EE
we obtain the inverse Laplace transformation (\ref{eq:inLap}) \cite{inverseLap}.

In order to compute the free energy by using the Matsubara method in the ideal gas approximation, we must compute the one-loop amplitude of strings. Much work has been done to generalize BSFT to the one-loop level of open strings on the {\D{p}} system \cite{1loop}. But there is an ambiguity in choosing the Weyl factors of the two boundaries of a one-loop world-sheet, because the conformal invariance is broken by the boundary terms in BSFT. Andreev and Oft have proposed the following form of 1-loop amplitude \cite{1loopAO} in type II string theory from the principle that its low energy part should coincide with that of the tachyon field model \cite{tachyon1} \cite{tachyon2}. If we restrict ourselves to the constant tachyon field, which we denote by $T$, it is given by
\BA
  {\it Z_{1}} &=& \frac{16 \pi^4 i \v}{(2 \pi \ap)^{\frac{p+1}{2}}}
    \int_{0}^{\infty} \frac{d \tau}{\tau}
      (4 \pi \tau)^{- \frac{p+1}{2}} e^{-4 \pi \T2 \tau} \no \\
  && \times \left[
    \left( \frac{\th_3 (0 | i \tau)}{{\th_1}' (0 | i \tau)} \right)^4
      - \left(\frac{\th_2 (0 | i \tau)}{{\th_1}' (0 | i \tau)} \right)^4 \right],
\EA
where $\v$ is the volume of the system that we are considering. Here we are considering only the coincident brane-antibrane system, and we will not treat the parallel brane and antibrane with a finite distance \cite{parallel} \cite{Huang2} in this paper. We can obtain the same amplitude when we consider the field theory which has the mass spectra
\BA
  {M_{NS}}^2
    &=& \frac{1}{\ap} \left( N_B + N_{NS} + 2 \T2 - \frac{1}{2} \right),
\label{eq:massNS} \\
  {M_{R}}^2
    &=& \frac{1}{\ap} \left( N_B + N_{R} + 2 \T2 \right),
\label{eq:massR}
\EA
where $M_{NS}$ and $M_{R}$ are the mass of the Neveu-Schwarz and Ramond sectors, respectively, and $N_B$, $N_{NS}$ and $N_{R}$ are the oscillation modes of the boson, Neveu-Schwarz fermion and Ramond fermion, respectively. If we use the proper time form of free energy \cite{Pol}, which is given by
\BA
  F (\B) &=& - \frac{\v}{(2 \pi \ap)^{\frac{p+1}{2}}}
    \int_{0}^{\infty} \frac{d \tau}{\tau}
      (4 \pi \tau)^{- \frac{p+1}{2}} \sum_{{M_{NS}}^2} \sum_{r=1}^{\infty}
        \exp \left( -2 \pi \ap {M_{NS}}^2 \tau
          - \frac{r^2 \B^2}{8 \pi \ap \tau} \right) \no \\
  && + \frac{\v}{(2 \pi \ap)^{\frac{p+1}{2}}}
    \int_{0}^{\infty} \frac{d \tau}{\tau}
      (4 \pi \tau)^{- \frac{p+1}{2}} \sum_{{M_{R}}^2} \sum_{r=1}^{\infty}
        (-1)^r \exp \left( -2 \pi \ap {M_{R}}^2 \tau
          - \frac{r^2 \B^2}{8 \pi \ap \tau} \right), \no \\
\EA
we can compute the free energy. The result is
\BA
  F (T, \B) &=& - \frac{16 \pi^4 \v}{(2 \pi \ap)^{\frac{p+1}{2}}}
    \int_{0}^{\infty} \frac{d \tau}{\tau}
      (4 \pi \tau)^{- \frac{p+1}{2}} e^{-4 \pi \T2 \tau} \no \\
  && \hspace{2cm} \times \left[ \left(\frac{\th_3 (0 | i \tau)}
    {{\th_1}' (0 | i \tau)} \right)^4
      \left( \th_3 \left( 0 \left| \frac{i \B^2}{8 \pi^2 \ap \tau} \right.
        \right) -1 \right) \right. \no \\
  && \hspace{4cm} - \left.
    \left( \frac{\th_2 (0 | i \tau)}{{\th_1}' (0 | i \tau)} \right)^4
      \left( \th_4 \left( 0 \left| \frac{i \B^2}{8 \pi^2 \ap \tau} \right.
        \right) -1 \right) \right].
\label{eq:freenoncompact}
\EA
This is the free energy on the non-compact background \cite{Hotta4}.

Let us generalize this free energy to that in the toroidal background. We consider the {\D{p}} system in type II string theory compactified on $D$-dimensional torus $T_{D}$. In the compact direction, there are two kinds of zero modes which give discrete energy spectra. They are momentum modes in the directions parallel to the {\D{p}} system, and winding modes in the directions transverse to it. If we include these modes to the mass spectra, it is given by
\BA
  {M_{NS}}^2
    &=& \sum_{I=1}^{p-d} \left( \frac{m_I}{R_I} \right)^2
      + \sum_{i=p-d+1}^{D} \left( \frac{n_i R_i}{\ap} \right)^2 
        + \frac{1}{\ap} \left( N_B + N_{NS} + 2 \T2 - \frac{1}{2} \right),
\label{eq:massNStorus} \\
  {M_{R}}^2
    &=& \sum_{I=1}^{p-d} \left( \frac{m_I}{R_I} \right)^2
      + \sum_{i=p-d+1}^{D} \left( \frac{n_i R_i}{\ap} \right)^2 
        + \frac{1}{\ap} \left( N_B + N_{R} + 2 \T2 \right).
\label{eq:massRtorus}
\EA
Here, we assumed that the {\D{p}} system is extended in the $d$-dimensional non-compact directions. It should be noted that these mass spectra are invariant under T-duality transformation \cite{Tdual}
\BE
  R_I \rightarrow \frac{\ap}{R_i} \ \ \ , \ \ \ m_I \rightarrow n_i,
\EE
for the directions parallel to the {\D{p}} system and
\BE
  R_i \rightarrow \frac{\ap}{R_I} \ \ \ , \ \ \ n_i \rightarrow m_I,
\EE
for the directions transverse to it. Summations over the momentum modes and winding modes can be rewritten by using $\th$ function as
\BA
  \sum_{m_I,n_i = - \infty}^{\infty}
    &\exp& \left[ -2 \pi \ap \tau \left(
      \sum_{I=1}^{p-d} \left( \frac{m_I}{R_I} \right)^2
        + \sum_{i=p-d+1}^{D} \left( \frac{n_i R_i}{\ap} \right)^2
          \right) \right] \no \\
  &=& \prod_{I=1}^{p-d} \prod_{i=p-d+1}^{D}
    \th_3 \left( 0 \left| \frac{2 i \ap \tau}{{R_I}^2} \right. \right)
      \th_3 \left( 0 \left| \frac{2 i {R_i}^2 \tau}{\ap} \right. \right).
\EA
Thus, the free energy is given by
\BA
  F (T, \B ,R) &=& - \frac{16 \pi^4 \vd}{{\BH}^{d+1}}
    \int_{0}^{\infty} \frac{d \tau}{\tau^{\frac{d+3}{2}}}
      e^{-4 \pi \T2 \tau} \prod_{I=1}^{p-d} \prod_{i=p-d+1}^{D}
        \th_3 \left( 0 \left| \frac{2 i \ap \tau}{{R_I}^2} \right. \right)
          \th_3 \left( 0 \left| \frac{2 i {R_i}^2 \tau}{\ap} \right. \right)
            \no \\
  && \hspace{2cm} \times \left[ \left(\frac{\th_3 (0 | i \tau)}
    {{\th_1}' (0 | i \tau)} \right)^4
      \left( \th_3 \left( 0 \left| \frac{i \B^2}{{\BH}^2 \tau} \right.
        \right) -1 \right) \right. \no \\
  && \hspace{4cm} - \left.
    \left( \frac{\th_2 (0 | i \tau)}{{\th_1}' (0 | i \tau)} \right)^4
      \left( \th_4 \left( 0 \left| \frac{i \B^2}{{\BH}^2 \tau} \right.
        \right) -1 \right) \right],
\label{eq:freecompact}
\EA
where $\vd$ is the $d$-dimensional volume in the non-compact directions parallel to the {\D{p}} system, and $\BH$ is the inverse of the Hagedorn temperature
\BE
  \BH = 2 \pi \sqrt{2 \ap}.
\EE
This free energy is invariant under the T-duality transformation. The finite temperature effective potential is also invariant under the T-duality transformation, because it is derived from this free energy. We only need to investigate the region $R \geq \sqrt{\ap}$, and we restrict ourselves to this region hereafter. In the next section, we will investigate the singular part of the free energy in the complex $\B$-plane.

\section{Complex Temperature Formalism}
\label{sec:comT}

Let us investigate the behavior of the free energy near singularities in the complex $\B$-plane. A general property of the free energy of strings is that it has the leading singularity at $\B = \BH$, which we call the Hagedorn singularity. If some radii are much larger than the string scale $\sqrt{\ap}$, there is the case that other radius dependent singularities exist close to the Hagedorn singularity \cite{Tan2} \cite{Hotta2} \cite{Thermo}. We will show these properties of the singular part of the free energy in this section.

In order to see the behavior of the free energy at high temperature, let us make the variable transformation
\BA
  \tau = \frac{1}{t}, \no
\EA
in (\ref{eq:freecompact}). Using the modular transformation of $\th$ functions, we obtain
\BA
  F (T, \B ,R) &=&
    - \frac{16 \pi^4 {\ap}^{d-p+ \frac{D}{2}} \vd}{2^{\frac{D}{2}} {\BH}^{d+1}}
      \left( \frac{\prod_{I=1}^{p-d} {R_I}^2}
        {\prod_{i=p-d+1}^{D} {R_i}^2} \right) \no \\
  && \hspace{1cm} \times \int_{0}^{\infty} dt \ t^{\frac{D+d-9}{2}}
    \exp \left( - \frac{4 \pi \T2}{t} \right)
      \prod_{I=1}^{p-d} \prod_{i=p-d+1}^{D}
        \th_3 \left( 0 \left| \frac{i {R_I}^2 t}{2 \ap} \right. \right)
          \th_3 \left( 0 \left| \frac{i \ap t}{2 {R_i}^2} \right. \right) \no \\
  && \hspace{2cm}
    \times \left[ \left(\frac{\th_3 (0 | i t)}{{\th_1}' (0 | it)} \right)^4
      \left( \th_3 \left( 0 \left| \frac{i \B^2 t}{{\BH}^2} \right. \right)
        -1 \right) \right. \no \\
  && \hspace{3cm} \left.
    - \left(\frac{\th_4 (0 | it)}{{\th_1}' (0 | it)} \right)^4
      \left( \th_4 \left( 0 \left| \frac{i \B^2 t}{{\BH}^2} \right. \right)
        -1 \right) \right].
\label{eq:Fmodular}
\EA
Expanding $\th$ functions and extracting the leading term in the large $t$ region near the Hagedorn singularity, we obtain
\BA
  F (T, \B ,R) &\simeq& - \frac{\A \vp}{\BH}
    \int_{\Lambda}^{\infty} dt
      \ t^{\frac{D+d-9}{2}} \exp \left[ - \frac{4 \pi \T2}{t}
        - \pi \frac{\B^2 - {\BH}^2}{{\BH}^2} t \right] \no \\
  && \times \prod_{I=1}^{p-d} \prod_{i=p-d+1}^{D}
    \th_3 \left( 0 \left| \frac{i {R_I}^2 t}{2 \ap} \right. \right)
      \th_3 \left( 0 \left| \frac{i \ap t}{2 {R_i}^2} \right. \right),
\label{eq:FHag}
\EA
where we have defined
\BE
  \A = \frac{{\ap}^{d-p+ \frac{D}{2}}}
    {2^{\frac{D}{2} -2} {\BH}^{d} \prod_{i=p-d+1}^{D} R_i},
\EE
and
\BE
  \vp = \vd \prod_{I=1}^{p-d} R_I,
\EE
and we have introduced the low energy cutoff $\Lambda$. This cutoff is required because we cannot apply above approximation to the low $\tau$ region. Since we will compute the $\T2$ term of the finite temperature effective potential in the vicinity of $T=0$, let us expand the free energy in $\T2$ and keep the lower order terms
\BA
  F (T, \B ,R) &\simeq& - \frac{\A \vp}{\BH}
    \int_{\Lambda}^{\infty} dt
      \ t^{\frac{D+d-9}{2}} \exp \left(
        - \pi \frac{\B^2 - {\BH}^2}{{\BH}^2} t \right) \no \\
  && \times \prod_{I=1}^{p-d} \prod_{i=p-d+1}^{D}
    \th_3 \left( 0 \left| \frac{i {R_I}^2 t}{2 \ap} \right. \right)
      \th_3 \left( 0 \left| \frac{i \ap t}{2 {R_i}^2} \right. \right) \no \\
  && + \frac{4 \pi \T2 \A \vp}{\BH}
    \int_{\Lambda}^{\infty} dt
      \ t^{\frac{D+d-11}{2}} \exp \left(
        - \pi \frac{\B^2 - {\BH}^2}{{\BH}^2} t \right) \no \\
  && \times \prod_{I=1}^{p-d} \prod_{i=p-d+1}^{D}
    \th_3 \left( 0 \left| \frac{i {R_I}^2 t}{2 \ap} \right. \right)
      \th_3 \left( 0 \left| \frac{i \ap t}{2 {R_i}^2} \right. \right).
\label{eq:FHag2}
\EA
We can evaluate the singular part of the free energy by computing the discontinuity around the singularity like in the previous paper \cite{Hotta4}. But let us compute it by using the incomplete $\Gamma$ function for the later convenience.

\subsection{String Scale Radius Case}
\label{sec:smallradiusF}

If all the radii are close to the string scale $\sqrt{\ap}$, the Hagedorn singularity is dominant and we only need to consider the terms
\BA
  F (T, \B ,R) &\simeq& - \frac{\A \vp}{\BH}
    \int_{\Lambda}^{\infty} dt
      \ t^{\frac{D+d-9}{2}} \exp \left(
        - \pi \frac{\B^2 - {\BH}^2}{{\BH}^2} t \right) \no \\
  && + \frac{4 \pi \T2 \A \vp}{\BH}
    \int_{\Lambda}^{\infty} dt
      \ t^{\frac{D+d-11}{2}} \exp \left(
        - \pi \frac{\B^2 - {\BH}^2}{{\BH}^2} t \right).
\label{eq:FHagRsmall}
\EA
This free energy is proportional to that in the non-compact background case \cite{Hotta4} if we replace $D+d$ by $p$. From this we can expect that a phase transition occurs only when $D+d=9$ and in fact it is true as we will see in the next section. We can rewrite (\ref{eq:FHagRsmall}) near $\B = \BH$ by using the incomplete $\Gamma$ function as
\BA
  F (T, \B ,R) &\simeq&
    - \frac{(2 \pi)^{\a} \A \vp (\B - \BH)^{\a}}{{\BH}^{\a +1}}
      \ \Gamma \left( - \a \ , \ 
        2 \pi \frac{\B - \BH}{\BH} \Lambda \right) \no \\
    && + \frac{2 (2 \pi)^{\a +2} \A \vp (\B - \BH)^{\a +1} \T2}{{\BH}^{\a +2}}
      \ \Gamma \left( - \a -1 \ , \ 
        2 \pi \frac{\B - \BH}{\BH} \Lambda \right),
\label{eq:Fcut2}
\EA
where we have defined
\BE
  \a \equiv \frac{7-D-d}{2}.
\EE
We can deduce the singular part of the free energy $F_{sing}$ from (\ref{eq:Fcut2}) as follows.

\renewcommand{\descriptionlabel}[1]{\large\bfseries{#1}}
\begin{description}

\item[(a)] {\large \bf \ $D+d=9$  ($\a = -1$)}
\vspace{0.5cm}

When $D+d=9$, which means $\a =-1$, the first argument of the incomplete $\Gamma$ function in the first term of (\ref{eq:Fcut2}) becomes one, so that we can set $\Lambda =0$. For the second term of (\ref{eq:Fcut2}) we can use the following formula for the incomplete $\Gamma$ function;
\BE
  \Gamma (0,x) = -\gamma - \log x
    - \sum_{n=1}^{\infty} \frac{(-1)^n x^n}{n \cdot n!},
\label{eq:Gammazero}
\EE
where $\gamma$ is the Euler constant. Combining these two terms, we get
\BE
  F_{sing} (T, \B ,R) \simeq - \frac{\A \vp}{2 \pi (\B - \BH)}
    - \frac{4 \pi \A \vp \T2}{\BH}
      \log \left( 2 \pi \frac{\B - \BH}{\BH} \Lambda \right).
\label{eq:FsingDd9}
\EE
We can take the same notation as the previous work \cite{Hotta4} by setting $\Lambda = (2 \pi)^{-1}$. We will set $\Lambda = (2 \pi)^{-1}$ in this formula hereafter.

\vspace{0.5cm}
\item[(b)] {\large \bf \ $D+d :$ even ($\a :$ half-integer)}
\vspace{0.5cm}

When $D+d$ is even, the first arguments of the incomplete $\Gamma$ functions are negative half-integers. In this case we can set $\Lambda =0$ and we obtain the singular part of the free energy as
\BA
  F_{sing} (T, \B ,R) &\simeq&
    - \frac{(2 \pi)^{\a} \Gamma ( - \a) \A \vp}{{\BH}^{\a +1}}
      \ (\B - \BH)^{\a} \no \\
    && + \frac{2 (2 \pi)^{\a +2} \Gamma ( - \a -1) \A \vp \T2}{{\BH}^{\a +2}}
      \ (\B - \BH)^{\a +1}.
\label{eq:FDdeven}
\EA

\vspace{0.5cm}
\item[(c)] {\large \bf \ $D+d :$ odd ($\a :$ integer)}
\vspace{0.5cm}

When $D+d$ is odd, the first arguments of the incomplete $\Gamma$ functions are negative integers. The incomplete $\Gamma$ function whose first argument is a negative integer can be expanded as
\BE
  \Gamma (-n,x) = \frac{1}{n!} e^{-x}
    \sum_{s=1}^{n} (-1)^{s-1} (n-s)! \ x^{-n+s-1}
      + \frac{(-1)^n}{n!} \Gamma (0,x).
\EE
In our case, $\Lambda \rightarrow 0$ corresponds to $x \rightarrow 0$, so that the incomplete $\Gamma$ function can be approximated by the last term. Thus, the singular part of the free energy can be obtained as
\BA
  F_{sing} (T, \B ,R) &\simeq&
    - \frac{(-1)^{\a +1} (2 \pi)^{\a} \A \vp}{\Gamma (\a +1) {\BH}^{\a +1}}
      \ (\B - \BH)^{\a}
        \log \left( 2 \pi \frac{\B - \BH}{\BH} \Lambda \right) \no \\
    && + \frac{(-1)^{\a +2} 2 (2 \pi)^{\a +2} \A \vp \T2}
      {\Gamma (\a +2) {\BH}^{\a +2}}
        \ (\B - \BH)^{\a +1}
          \log \left( 2 \pi \frac{\B - \BH}{\BH} \Lambda \right). \no \\
\label{eq:FDdodd}
\EA
We will also set $\Lambda = (2 \pi)^{-1}$ in this formula hereafter.

\end{description}

\subsection{Large Radius Case}
\label{sec:LargeradiusF}

Let us consider the case that some radii are much larger than the string scale $\sqrt{\ap}$. When the radii in the directions parallel to the {\D{p}} system are much larger than the string scale, the Hagedorn singularity is dominant and we can use the results above. But when the radii in the directions transverse to the {\D{p}} system are much larger than the string scale, the radius dependent singularities exist near the Hagedorn singularity as is sketched in Figure 1, and we must consider the radius dependent terms. We will show only the $D+d=9$ case, because, in the $D+d \leq 8$ case, the result can be obtained in a similar way to the $D+d=9$ case as we will comment in the next section. For simplicity, let us assume that some of the transverse directions has the same radius $R_{\db}$ with $R_{\db} \gg \sqrt{\ap}$. Since $\th$ function in (\ref{eq:FHag2}) can be rewritten as
\begin{figure}
\begin{center}
$${\epsfxsize=6.5 truecm \epsfbox{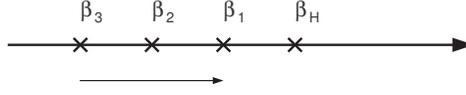}}$$
\caption{Radius dependent singularities}
\label{fig:complex1}
\end{center}
\end{figure}
\BE
  \prod_{i=p-d+1}^{p-d+ \db}
    \th_3 \left( 0 \left| \frac{i \ap t}{2 {R_{\db}}^2} \right. \right)
      = \sum_{n_i =- \infty}^{\infty} \exp \left( - \sum_{i=p-d+1}^{p-d+ \db}
        \frac{\pi {n_i}^2 \ap}{2 {R_{\db}}^2} t \right),
\EE
the free energy can be approximated as
\BA
  F (T, \B ,R) &\simeq& - \frac{\A \vp}{\BH}
    \int_{\Lambda}^{\infty} dt
      \sum_{n_i =- \infty}^{\infty} \exp \left(
        - \pi \frac{\B^2 - {\Bn}^2}{{\BH}^2} t \right) \no \\
  && + \frac{4 \pi \T2 \A \vp}{\BH}
    \int_{\Lambda}^{\infty} \frac{dt}{t}
      \sum_{n_i =- \infty}^{\infty} \exp \left(
        - \pi \frac{\B^2 - {\Bn}^2}{{\BH}^2} t \right),
\label{eq:FHagRlarge}
\EA
where we have defined
\BE
  {\Bn}^2 \equiv {\BH}^2 \left( 1- \sum_{i=p-d+1}^{p-d+ \db}
    \frac{{n_i}^2 \ap}{2 {R_{\db}}^2} \right).
\label{eq:Bn}
\EE
From this we can see that singularities exist at $\B = \Bn$ on the real axis of complex $\B$-plane, and they approach to the Hagedorn singularity $\B = \BH$ as $R_{\db}$ increases. This free energy can be rewritten by using the incomplete $\Gamma$ function as
\BA
  F (T, \B ,R) &\simeq&
    \sum_{n_i =- \infty}^{\infty} \left[
      - \frac{\BH \A \vp}{2 \pi \Bn (\B - \Bn)}
      \ \Gamma \left( 1 \ , \ 
        2 \pi \frac{\Bn (\B - \Bn)}{{\BH}^2} \Lambda \right) \right. \no \\
    && \hspace{2cm} \left. + \frac{4 \pi \A \vp \T2}{\BH}
      \ \Gamma \left( 0 \ , \ 
        2 \pi \frac{\Bn (\B - \Bn)}{{\BH}^2} \Lambda \right) \right],
\label{eq:FGammalarge}
\EA
where we assumed that $\B \simeq \Bn$ for each terms. For the first term, we can set $\Lambda =0$. For the second term, we can use (\ref{eq:Gammazero}). Combining these two terms, we get
\BE
  F_{sing} (T, \B ,R) \simeq \sum_{n_i =- \infty}^{\infty} \left[
    - \frac{\BH \A \vp}{2 \pi \Bn (\B - \Bn)}
      - \frac{4 \pi \A \vp \T2}{\BH}
        \log \left( 2 \pi \frac{\Bn (\B - \Bn)}{{\BH}^2} \Lambda \right) \right].
\label{eq:Fsinglarge}
\EE
We will also set $\Lambda = (2 \pi)^{-1}$ in this formula hereafter.

\section{Finite Temperature Effective Potential near the Hagedorn Temperature}
\label{sec:poten}

In this section, we compute the finite temperature effective potential of the brane-antibrane systems in the toroidal background. We must compute it by using the microcanonical ensemble method as we have mentioned in \S \ref{sec:free}. All the statistical variables are derived from the density of states in the microcanonical ensemble method. The density of states $\Omega (E)$ is obtained from the inverse Laplace transformation of the partition function $Z( \B )$ (\ref{eq:partitionfree}). The partition function is given by (\ref{eq:partitionfree}), which is rewritten as
\BE
  Z(\B) = \exp \left[ - \B (F_{reg} + F_{sing}) \right],
\label{eq:ZeW}
\EE
where $F_{reg}$ is the regular part of the free energy. The entropy $S(E)$ can be derived from $\Omega (E)$ as
\BE
  S(E) = \log \Omega (E) \delta E,
\label{eq:Sdef}
\EE
where $\delta E$ represents the energy fluctuation, and the inverse temperature $\B$ from the partial derivative of S(E)
\BE
  \B = \frac{\p S}{\p E}.
\label{eq:Bdef}
\EE
The finite temperature effective potential of the {\D{p}} system can be calculated from these variables as
\BE
  V(T,E) = 2 \tau_{p} \vp \exp (-8 \T2) - \B^{-1} S.
\label{eq:Vdef}
\EE
Now, we are ready to calculate the finite temperature effective potential by using the microcanonical ensemble method.

\subsection{String Scale Radius Case}
\label{sec:smallradius}

Let us first consider the case that all the radii in the transverse directions are about string scale. In this case, the calculation of the finite temperature effective potential in the toroidal background is similar to that on non-compact background. Thus, we will calculate in the $D+d=9$ and the $D+d=8$ cases explicitly, and show only the results for the other cases.

\renewcommand{\descriptionlabel}[1]{\large\bfseries{#1}}
\begin{description}

\item[(a)] {\large \bf \ $D+d=9$  ($\a = -1$)}
\vspace{0.5cm}

Let us consider the $D+d=9$ case, which is the most interesting case. This is the case that the {\D{p}} system is extended in all the non-compact directions. In this case, $F_{sing}$ is given by (\ref{eq:FsingDd9}) and $F_{reg}$ can be expanded as
\BE
  - \B F_{reg} (\B) = \lambda_0 \vp - \sigma_0 \vp (\B - \BH)
    + O ( \vp (\B - \BH)^2 ).
\label{eq:Fregexp1}
\EE
Then, by using (\ref{eq:FsingDd9}) the inverse Laplace transformation (\ref{eq:inLap}) can be rewritten as
\BE
  \Omega (T,E,R) \simeq e^{\BH E + \lambda_0 \vp} \int_{C_a} \frac{d \B}{2 \pi i}
    \left( \frac{\B - \BH}{\BH} \right)^{- \Dm \vp \T2}
      \exp \left[ (\B - \BH) \Eb + \frac{\Cm \vp}{\B - \BH} \right],
\label{eq:Dd9dos}
\EE
where we have defined
\BA
  \Cm &=& \frac{\BH \A}{2 \pi}, 
    \label{eq:Cm} \\
  \Dm &=& -4 \pi \A,
\EA
and we take the contour $C_a$ as sketched in Figure 2.
\begin{figure}
\begin{center}
$${\epsfxsize=6.5 truecm \epsfbox{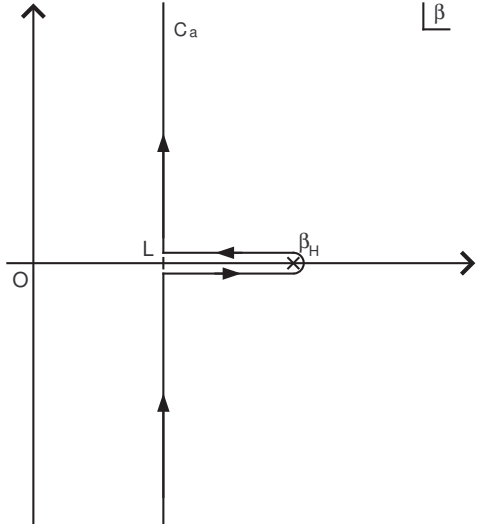}}$$
\caption{Complex $\B$-plane}
\label{fig:complex2}
\end{center}
\end{figure}
Let us suppose that both $E$ and $E / \vp$ are very large. Then, the saddle point method works well because the exponent in the integrand is very large. The result is
\BE
  \Omega (T,E,R) \simeq \frac{1}{2 \sqrt{\pi}}
    \left( \frac{{\BH}^2 \Eb}{\Cm \vp}
      \right)^{\frac{1}{2} \Dm \vp \T2}
        \left( \frac{\Cm \vp}{\Eb^3} \right)^{\frac{1}{4}}
          \exp \left( \BH E + \lambda_0 \vp
            +2 \sqrt{\Cm \vp \Eb} \right).
\EE
Substituting $\Omega (E)$ into (\ref{eq:Sdef}), we obtain
\BA
  S(T,E,R) \simeq
    \frac{\Dm \vp \T2}{2} \log \left( \frac{{\BH}^2 \Eb}{\Cm \vp} \right)
      - \frac{3}{4} \log \left( \frac{\Eb}{{\Cm}^{\frac{1}{3}}
        \vp^{\frac{1}{3}} (\delta E)^{\frac{4}{3}}} \right) \no \\
          + \BH E + \lambda_0 \vp + 2 \sqrt{\Cm \vp \Eb},
\EA
and from (\ref{eq:Bdef}), we get
\BE
  \B \simeq \frac{\Dm \vp \T2}{2 \Eb}
    - \frac{3}{4 \Eb} + \BH + \sqrt{\frac{\Cm \vp}{\Eb}}.
\label{eq:Bmin}
\EE
The finite temperature effective potential can be derived from (\ref{eq:Vdef}). In order to argue the stability of the brane-antibrane system, we need only $\T2$ term of $V(T,E,R)$. This term is given by
\BE
  \left[ -16 \tau_p \vp
   + \frac{2 \pi \A \vp}{\BH}
    \log \left( \frac{2 \pi \BH \Eb}{\A \vp}
     \right) \right] \T2.
\label{eq:p9T2E}
\EE
It should be noted that the second term in the coefficient of $\T2$ increases with increasing $E$. Since the first term is constant as far as $\vp$ and $\tau_p$ fixed, the sign of the $\T2$ term changes from negative to positive at large $E$. The coefficient vanishes when
\BE
  \Eb \simeq \frac{\A \vp}{2 \pi \BH}
    \exp \left( \frac{8 \BH \tau_p}{\pi \A} \right).
\label{eq:EDd9}
\EE
Since (\ref{eq:Bmin}) can be approximated as
\BE
  \B \simeq \BH + \sqrt{\frac{\Cm \vp}{\Eb}},
\EE
for large $E$, we can derive the critical temperature ${\cal T}_c$ at which the coefficient vanishes. The result is
\BE
  {\cal T}_c = \beta^{-1}
   \simeq {\BH}^{-1}
    \left[ 1+ \exp \left( - \frac{4 \BH \tau_p}
     {\pi \A} \right) \right]^{-1}.
\label{eq:Tc}
\EE
Here we see that this temperature is very close to the Hagedorn temperature since $\tau_p$ is very large if the coupling of strings is very small. Above this temperature, the coefficient of $\T2$ is positive and $T=0$ becomes the potential minimum. This implies that a phase transition occurs at the temperature ${\cal T}_c$ which is slightly below the Hagedorn temperature, and the {\D{p}} system, which is extended in all the non-compact directions, is stable above this temperature.

\vspace{0.5cm}
\item[(b)] {\large \bf \ $D+d=8$  ($\a = - 1/2$)}
\vspace{0.5cm}

Let us consider the $D+d=8$ case, where we will see that a result strikingly different from the $D+d=9$ case arises. This is the case that the {\D{p}} system is extended in all the non-compact directions except for one direction. In this case, the inverse Laplace transformation (\ref{eq:inLap}) is rewritten as
\BA
  \Omega (T,E,R) &\simeq& e^{\BH E + \lambda_0 \vp}
    \int_{C_a} \frac{d \B}{2 \pi i}
      \exp \left[ (\B - \BH) \Eb
        + \Ch \vp (\B - \BH)^{- \frac{1}{2}} \right. \no \\
  && \hspace{5.5cm} \left. - \Dh \vp \T2 (\B - \BH)^{\frac{1}{2}} \right] \no \\
  &\simeq& e^{\BH E + \lambda_0 \vp} \int_{C_a} \frac{d \B}{2 \pi i}
    \left[ 1- \Dh \vp \T2 (\B - \BH)^{\frac{1}{2}} \right] \no \\
  && \hspace{3cm} \times \exp \left[ (\B - \BH) \Eb
    + \Ch \vp (\B - \BH)^{- \frac{1}{2}} \right],
\EA
where we take the small $\T2$ approximation in the second equality and we have defined
\BA
  \Ch &=& \frac{{\BH}^{\frac{1}{2}} \A}{\sqrt{2}},  \\
  \Dh &=& \frac{2^{\frac{7}{2}} \pi^2 \A}{{\BH}^{\frac{1}{2}}}.
\EA
We can also use the saddle point method and obtain
\BE
  \Omega (T,E,R) \simeq \frac{{\Ch}^{\frac{1}{3}} {\vp}^{\frac{1}{3}}}
    {3^{\frac{1}{2}} 2^{\frac{1}{3}} \pi^{\frac{1}{2}} \Eb^{\frac{5}{6}}}
      \exp \left( \BH E + \lambda_0 \vp
        + \frac{3 {\Ch}^{\frac{2}{3}} \vp^{\frac{2}{3}} \Eb^{\frac{1}{3}}}
          {2^{\frac{2}{3}}}
            - \frac{{\Ch}^{\frac{1}{3}} \Dh \vp^{\frac{4}{3}} \T2}
              {2^{\frac{1}{3}} \Eb^{\frac{1}{3}}} \right).
\EE
We can calculate the entropy $S$, the inverse temperature $\B$ and the potential $V(T,E,R)$ from (\ref{eq:Sdef}), (\ref{eq:Bdef}) and (\ref{eq:Vdef}) as in the $D+d=9$ case. The $\T2$ term of $V(T,E,R)$ is given by
\BE
  \left[ -16 \tau_p \vp
   - \frac{2^5 \pi^2 {\A}^{\frac{4}{3}} \vp^{\frac{4}{3}}}
     {3 {\BH}^{\frac{4}{3}} \Eb^{\frac{1}{3}}}
     \right] \T2.
\label{eq:T2p8}
\EE
It should be noted that the second term in the coefficient of $\T2$ decreases as $E$ gets large. Thus, the coefficient of $\T2$ remains negative for large $E$. This implies that a phase transition does not occur unlike in the $D+d=9$ case.

\vspace{0.5cm}
\item[(c)] {\large \bf \ $D+d \leq 7$ ($\a \leq 0$)}
\vspace{0.5cm}

We only show the results for the $D+d \leq 7$ case, because we can obtain these results by the similar calculation to the previous work \cite{Hotta4}. The $\T2$ term of the finite temperature effective potential is given by
\BE
  \left[ -16 \tau_p \vp
   - \frac{8 \pi^2 {\A}^2 \vp^2}{\BH \Eb}
     \log \left( \frac{\BH \Eb}{\A \vp} \right) \right] \T2,
\label{eq:T2p7}
\EE
for $D+d = 7$,
\BE
  \left[ -16 \tau_p \vp
   + \frac{2^{10} \pi^6 {\A}^4 \vp^4}{6 {\BH}^4 \Eb^3} \right] \T2,
\EE
for $D+d = 6$,
\BE
  \left[ -16 \tau_p \vp
   + \frac{16 \pi^2}{{\BH}^{2} E'} \right] \T2,
\EE
for $D+d = 5$, where
\BE
  E' \equiv \Eb - \frac{2 \pi \A}{\BH} \vp \log \left( \A \vp \right),
\EE
and
\BE
  \left[ -16 \tau_p \vp
   + \frac{16 \pi^2}{{\BH}^{2} \Eb} \right] \T2,
\EE
for $D+d \leq 4$. From these we can see that the coefficients are negative for large $E$, so that a phase transition does not occur.

\end{description}

\subsection{Large Radius Case}
\label{sec:Largeradius}

Next, let us consider the case that some of the radii in the directions transverse to the {\D{p}} system are much larger than the string scale. In this case, the radius dependent singularities exist near the Hagedorn singularity. The behavior of the finite temperature effective potential is determined by the relation between the total energy and the radii of the target space torus. We will mainly consider the $D+d=9$ case. We assume that all the $\db$-dimensional directions have the same radius $R_{\db}$, for simplicity.

\renewcommand{\descriptionlabel}[1]{\large\bfseries{#1}}
\begin{description}

\item[(a)] {\large \bf \ Hagedorn Singularity Dominance, Saddle Point Method}
\vspace{0.5cm}

In order to estimate the contribution from the radius dependent singularities, let us pick up the leading two singular terms of the free energy (\ref{eq:Fsinglarge}). Then the inverse Laplace transformation (\ref{eq:inLap}) can be rewritten as
\BE
  \Omega (T,E,R) \simeq \int_{C_a} \frac{d \B}{2 \pi i}
    \exp \left[ \B E - \B F_{reg} (\B) + \frac{\Cm \vp}{\B - \BH}
      + 2 \db \ \frac{\Cm \vp}{\B - {\B}_{1}} \right],
\EE
where $\Cm$ is defined by (\ref{eq:Cm}) and $\B_{1}$ is the nearest singularity to the Hagedorn singularity, namely, $\Bn$ with $\sum_i {n_i}^2 =1$. In this case, we can deform the contour as sketched in Figure 3, and we obtain
\begin{figure}
\begin{center}
$${\epsfxsize=6.5 truecm \epsfbox{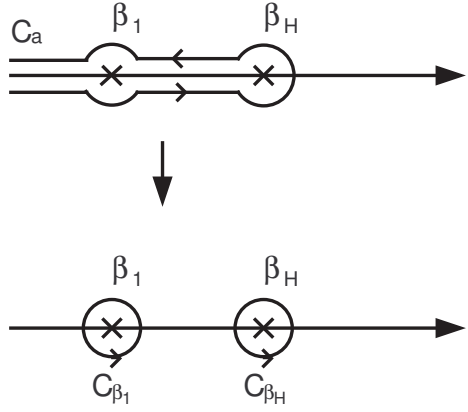}}$$
\caption{Contour deformation}
\label{fig:complex3}
\end{center}
\end{figure}
\BA
  \Omega (T,E,R) &\simeq& \int_{C_{\BH}} \frac{d \B}{2 \pi i}
    \exp \left[ \B E - \B F_{reg} (\B) + \frac{\Cm \vp}{\B - \BH} \right] \no \\
  && + \int_{C_{\B_{1}}} \frac{d \B}{2 \pi i}
    \exp \left[ \B E - \B F_{reg} (\B)
      + 2 \db \ \frac{\Cm \vp}{\B - {\B}_{1}} \right].
\label{eq:Doslarge1}
\EA
Let us expand $F_{reg} (\B)$ in a power of $(\B - \BH)$ as (\ref{eq:Fregexp1}) and of $(\B - \B_{1})$ as
\BE
  - \B F_{reg} (\B) = \lambda_1 \vp - \sigma_1 \vp (\B - \B_{1})
    + O ( \vp (\B - \B_{1})^2 ),
\label{eq:Fregexp2}
\EE
and define
\BA
  {\Eb}' &\equiv& E- \sigma_1 \vp, \no \\
  z &\equiv& \sqrt{\frac{\Eb}{\Cm \vp}} \ (\B - \BH), \no \\
  z' &\equiv& \sqrt{\frac{{\Eb}'}{2 \db \Cm \vp}} \ (\B - \B_{1}). \no \\
\EA
Then (\ref{eq:Doslarge1}) can be rewritten as
\BA
  \Omega (T,E,R) &\simeq& \sqrt{\frac{\Cm \vp}{\Eb}} e^{\BH E + \lambda_0 \vp}
    \oint \frac{dz}{2 \pi i}
      \exp \left[ \sqrt{\Cm \vp \Eb} \left(z+ \frac{1}{z} \right) \right] \no \\
  && + \sqrt{\frac{2 \db \Cm \vp}{{\Eb}'}} e^{\B_{1} E + \lambda_1 \vp}
    \oint \frac{dz'}{2 \pi i}
      \exp \left[ \sqrt{2 \db \Cm \vp {\Eb}'}
        \left(z'+ \frac{1}{z'} \right) \right], \no \\
\label{eq:Doslarge2}
\EA
where the contours for each terms encircle the origin in the complex $z$-plane and $z'$-plane counter-clockwise respectively. If we assume that $\sqrt{\Cm \vp \Eb} \gg 1$, then we can use the saddle point method. Substituting (\ref{eq:Cm}), we can rewrite this condition as
\BE
  \Eb \gg \frac{{\ap}^{\frac{p- \db-1}{2}} {R_{\db}}^{\db}}{\vp}.
\EE
By using the saddle point method, we obtain
\BE
  \Omega (T,E,R) \simeq \Omega_0 (T,E,R) + \Omega_1 (T,E,R),
\EE
where $\Omega_0 (T,E,R)$ and $\Omega_1 (T,E,R)$ represent the contribution from the Hagedorn singularity $\B = \BH$ and from the next leading singularity $\B = \B_1$ respectively:
\BA
  \Omega_0 (T,E,R) &\simeq& \frac{1}{2 \sqrt{\pi}}
    \left( \frac{\Cm \vp}{\Eb^3} \right)^{\frac{1}{4}}
      \exp \left( \BH E + \lambda_0 \vp
        +2 \sqrt{\Cm \vp \Eb} \right), \no \\
  \Omega_1 (T,E,R) &\simeq& \frac{{\db}^{\frac{1}{4}}}{2^{\frac{3}{4}}
    \sqrt{\pi}} \left( \frac{\Cm \vp}{{{\Eb}'}^3} \right)^{\frac{1}{4}}
      \exp \left( \B_{1} E + \lambda_1 \vp
        +2 \sqrt{2 \db \Cm \vp {\Eb}'} \right). \no
\EA
If $\log (\Omega_0 / \Omega_1) \gg 1$, then $\Omega (T,E,R)$ is dominated by $\Omega_0 (T,E,R)$ and we can ignore the term $\Omega_1 (T,E,R)$ \cite{Tan2} \cite{Thermo}. Let us assume that $E \gg \sigma_0 \vp$. Then $\log (\Omega_0 / \Omega_1)$ can be approximated as
\BA
  \log \frac{\Omega_0}{\Omega_1}
    &\simeq& \left( \BH E + \lambda_0 \vp +2 \sqrt{\Cm \vp \Eb} \right)
      - \left( \B_{1} E + \lambda_1 \vp
        +2 \sqrt{2 \db \Cm \vp {\Eb}'} \right) \no \\
  &\simeq& ( \BH - \B_{1} ) \Eb
    - \left( \sqrt{2 \db} -1 \right) 2 \sqrt{\Cm \vp \Eb},
\EA
where, in the second equality, we have used $(\lambda_0 - \lambda_1) \vp \simeq - \sigma_0 \vp ( \BH - \B_{1} )$, which is derived from (\ref{eq:Fregexp1}) and (\ref{eq:Fregexp2}). Then above condition is approximated as
\BE
  ( \BH - \B_{1} ) \Eb \gg \sqrt{\Cm \vp \Eb}.
\EE
Substituting (\ref{eq:Cm}) and (\ref{eq:Bn}) with $\sum_i {n_i}^2 =1$, we obtain
\BE
  \Eb \gg \frac{{R_{\db}}^{4- \db} \vp}{{\ap}^{\frac{5+p- \db}{2}}}.
\EE
We can also show that we can ignore the contribution from other singularities. Therefore, if $\Eb$ satisfies
\BE
  \Eb \gg \max \left( \sigma_0 \vp,
    \frac{{\ap}^{\frac{p- \db-1}{2}} {R_{\db}}^{\db}}{\vp} ,
      \frac{{R_{\db}}^{4- \db} \vp}{{\ap}^{\frac{5+p- \db}{2}}} \right),
\label{eq:regionA}
\EE
then the density of states is dominated by the contribution from the Hagedorn singularity and we can use the saddle point method to compute the density of states. By using the saddle point method, we can also show the Hagedorn singularity dominance in $\T2$ terms. As a consequence, $V(T,E,R)$ is the same as that in the case of $\db =0$ and $D+d=9$, and phase transition occurs near the Hagedorn temperature.

\vspace{0.5cm}
\item[(b)] {\large \bf \ Hagedorn Singularity Dominance, No Saddle Point Method}
\vspace{0.5cm}

Let us consider the region where we cannot use the saddle point method. From (\ref{eq:Doslarge2}) the saddle point approximation does not work well when $\sqrt{\Cm \vp \Eb} \ll 1$, namely,
\BE
  \Eb \ll \frac{{\ap}^{\frac{p- \db-1}{2}} {R_{\db}}^{\db}}{\vp}.
\label{eq:ERdd}
\EE
In this case, the exponents in the integrands in (\ref{eq:Doslarge2}) are very small and both integrals are given by powers of $\sqrt{\Cm \vp \Eb}$. Thus, $\log (\Omega_0 / \Omega_1)$ can be approximated as
\BE
  \log \frac{\Omega_0}{\Omega_1} \simeq ( \BH - \B_{1} ) \Eb,
\EE
we can ignore the term $\Omega_1 (T,E,R)$ if $\Eb$ satisfies $( \BH - \B_{1} ) \Eb \gg 1$, which is rewritten as
\BE
  \Eb \gg \frac{{R_{\db}}^2}{{\ap}^{\frac{3}{2}}}.
\EE
We can also show that we can ignore the contribution from other singularities in this region. Therefore, if $\Eb$ satisfies
\BE
  \max \left( \sigma_0 \vp , \frac{{R_{\db}}^2}{{\ap}^{\frac{3}{2}}} \right)
    \ll \Eb \ll \frac{{\ap}^{\frac{p- \db-1}{2}} {R_{\db}}^{\db}}{\vp},
\label{eq:regionB}
\EE
then the density of states is dominated by the contribution from the Hagedorn singularity and we can ignore the contribution from other singularities. Thus, the density of states is given by (\ref{eq:Dd9dos}), which can be rewritten by using the modified Bessel function of the first kind as \cite{Thermo}
\BE
  \Omega (T,E,R) \simeq 
    \left( \frac{\Cm \vp}{\Eb} \right)^{\frac{1- \Dm \vp \T2}{2}}
      {\BH}^{\Dm \vp \T2} e^{\BH E + \lambda_0 \vp} \
        I_{\Dm \vp \T2 -1} \left( 2 \sqrt{\Cm \vp \Eb} \right).
\EE
Let us expand the modified Bessel function as
\BE
  I_{\nu} (z) = \sum_{n=0}^{\infty} \frac{1}
    {n! \ \Gamma (\nu +n+1)} \left( \frac{z}{2} \right)^{2n+ \nu},
\EE
and keep the lower order terms because the argument $2 \sqrt{\Cm \vp \Eb} \ll 1$ is very small. Using the fact that the $\Gamma$ function can be approximated as
\BE
  \Gamma (\epsilon) \simeq \frac{1}{\epsilon},
\EE
for small $\epsilon$, we obtain
\BE
  \Omega (T,E,R) \simeq \Cm \vp (\BH \Eb)^{\Dm \vp \T2}
    e^{\BH E + \lambda_0 \vp}
      \left( 1+ \frac{\Dm \T2}{\Cm \Eb} \right).
\EE
We can calculate the entropy $S$, the inverse temperature $\B$ and the potential $V(T,E,R)$ from (\ref{eq:Sdef}), (\ref{eq:Bdef}) and (\ref{eq:Vdef}) as in the string scale radius case. The $\T2$ term of $V(T,E,R)$ is given by
\BE
  \left[ -16 \tau_p \vp
   + \frac{4 \pi \A \vp}{\BH}
    \log ( \BH \Eb ) \right] \T2.
\label{eq:RlargeT2E}
\EE
The coefficient vanishes when
\BE
  \Eb \simeq \frac{1}{\BH}
    \exp \left( \frac{2^{\frac{9-d}{2}} {\BH}^{d+1} \tau_p {R_{\db}}^{\db}}
      {\pi {\ap}^{\frac{d+ \db -p}{2}}} \right),
\EE
where we use the explicit expression of $\A$ with $D+d=9$, that is,
\BE
  \A = \frac{{\ap}^{\frac{d+ \db -p}{2}}}
    {2^{\frac{5-d}{2}} {\BH}^d {R_{\db}}^{\db}}.
\EE
But this $\Eb$ cannot satisfy the condition (\ref{eq:ERdd}). This is because $\tau_{p}$ is very large if the coupling of strings is very small, and $\vp \geq {\ap}^{\frac{p}{2}}$ since we assumed that $R \geq \sqrt{\ap}$. This means that, as $E$ increases, $\Eb$ exceeds the region (\ref{eq:regionB}) before a phase transition occurs. Thus, a phase transition does not occur in this region.

\vspace{0.5cm}
\item[(c)] {\large \bf \ Contribution from Radius Dependent Singularities}
\vspace{0.5cm}

In the previous two regions (\ref{eq:regionA}) and (\ref{eq:regionB}), the density of states is dominated by the contribution from the Hagedorn singularity. Here, we consider the remaining region
\BE
  \sigma_0 \vp \ll \Eb \ll \min \left( \frac{{R_{\db}}^2}{{\ap}^{\frac{3}{2}}} ,
    \frac{{\ap}^{\frac{p- \db-1}{2}} {R_{\db}}^{\db}}{\vp} \right),
\EE
or
\BE
  \max \left( \sigma_0 \vp ,
    \frac{{\ap}^{\frac{p- \db-1}{2}} {R_{\db}}^{\db}}{\vp} \right)
      \ll \Eb \ll
        \frac{{R_{\db}}^{4- \db} \vp}{{\ap}^{\frac{5+p- \db}{2}}},
\EE
in which we cannot ignore the contribution from the radius dependent singularities. The singular part of the free energy $F_{sing}$ is given by (\ref{eq:Fsinglarge}). Although we must perform the summation over $n_i$ in order to include the contribution from the radius dependent singularities, it is difficult to compute this sum directly. However if we return to (\ref{eq:FHagRlarge}), which is rewritten as
\BA
  F (T, \B ,R) &\simeq& - \frac{\A \vp}{\BH}
    \int_{\Lambda}^{\infty} dt
      \sum_{n_i =- \infty}^{\infty}
        \exp \left[ - \sum_{i=p-d+1}^{p-d+ \db}
          \frac{\pi {n_i}^2 \ap}{2 {R_{\db}}^2} t
            - \pi \frac{\B^2 - {\BH}^2}{{\BH}^2} t \right] \no \\
  && + \frac{4 \pi \T2 \A \vp}{\BH}
    \int_{\Lambda}^{\infty} \frac{dt}{t}
      \sum_{n_i =- \infty}^{\infty}
        \exp \left[ - \sum_{i=p-d+1}^{p-d+ \db}
          \frac{\pi {n_i}^2 \ap}{2 {R_{\db}}^2} t
            - \pi \frac{\B^2 - {\BH}^2}{{\BH}^2} t \right], \no \\
\EA
the sum over $n_i$ is approximated by integral as
\BE
  \prod_{i=p-d+1}^{p-d+ \db} \sum_{n_i =- \infty}^{\infty}
    \exp \left( - \frac{\pi {n_i}^2 \ap}{2 {R_{\db}}^2} t \right)
      \longrightarrow \prod_{i=p-d+1}^{p-d+ \db}
        \left( \frac{2 {R_{\db}}^2}{\pi \ap t} \right)^{\frac{1}{2}}
          \int_{- \infty}^{\infty} e^{- {x_i}^2} d x_i,
\EE
for large $R_{\db}$, and we can compute it. By evaluating the gaussian integral, we obtain
\BA
  F (T, \B ,R) &\simeq& - \frac{\Ap \vp}{\BH}
    \int_{\Lambda}^{\infty} dt \ t^{- \frac{\db}{2}}
      \exp \left( - \pi \frac{\B^2 - {\BH}^2}{{\BH}^2} t \right) \no \\
  && + \frac{4 \pi \T2 \Ap \vp}{\BH}
    \int_{\Lambda}^{\infty} dt \ t^{- \frac{\db +2}{2}}
      \exp \left( - \pi \frac{\B^2 - {\BH}^2}{{\BH}^2} t \right), \no \\
\EA
where we have defined
\BE
  \Ap = \frac{{\ap}^{-p+ \frac{9+d- \db}{2}}}
    {2^{\frac{5-d- \db}{2}} {\BH}^d \prod_{i=p-d+ \db +1}^{9-d} R_i}.
\EE
This free energy is equivalent to (\ref{eq:FHagRsmall}) with $D+d=9- \db$. If we compute the finite temperature effective potential from this, we obtain the same results with the string scale radius case with $D+d=9- \db$. As a consequence, a phase transition does not occur in this region.

\end{description}

Here, let us make a short comment on the $D+d \leq 8$ case. For extremely large $E$, we can show that the contribution from the Hagedorn singularity is dominant, and we obtain the same results with the string scale radius case with the same $D+d$. For not so large $E$, the summation over $n_i$ can be approximated by the gaussian integral, and we obtain the same results with the string scale radius case with $D+d$ replaced by $D+d- \db$. Therefore, a phase transition does not occur in this case.

\section{Conclusion and Discussion}
\label{sec:conclusion}

In this paper, we have discussed the behavior of the finite temperature effective potential on the brane-antibrane pair in the toroidal background. We have evaluated the $\T2$ term of the potential near the Hagedorn temperature by using the microcanonical ensemble method, and investigated whether $T=0$ becomes stable near the Hagedorn temperature.

When all the radii in the directions transverse to the {\D{p}} system are about the string scale, the result of the $D+d=9$ case is in sharp contrast to that of the $D+d \leq 8$ case. For $D+d=9$ case, the sign of the coefficient of the $\T2$ term changes from negative to positive at slightly below the Hagedorn temperature. This implies that a phase transition occurs at this temperature and the {\D{p}} system becomes stable above this temperature. For $D+d \leq 8$ case, the coefficient remains negative near the Hagedorn temperature, so that such a phase transition does not occur. We thus concluded that the {\D{p}} pair, which is extended in all the non-compact directions, is created near the Hagedorn temperature.

When some of radii in the directions transverse to the {\D{p}} system are much larger than the string scale, the result of the $D+d=9$ case is also in sharp contrast to that of the $D+d \leq 8$ case. In $D+d=9$ case, a phase transition occurs if $\Eb$ reaches (\ref{eq:EDd9}) which satisfies (\ref{eq:regionA}). Therefore, a phase transition occurs for large enough $E$ even if $R_{\db}$ is much larger than the string scale and the {\D{p}} system is not extended in all the $\db$-directions. Whether phase transition occurs or not depends on the topology of the target space. Strings are sensitive to the topology of the target space. In $D+d \leq 8$ case, there is no possibility that a phase transition occurs.

Since our calculation is based on the work of Andreev and Oft \cite{1loopAO}, we must confirm the validity of their work. However, we only need to analyze the vicinity of $T=0$ in order to investigate whether the phase transition occurs or not. Our calculation is valid if the mass square is shifted as equations (\ref{eq:massNStorus}) and (\ref{eq:massRtorus}) for small $|T|$.

The conclusion of our previous work is that only the spacetime-filling {\D{9}} pair is created near the Hagedorn temperature in the non-compact background \cite{Hotta4}. Since the {\D{9}} pair always satisfies $D+d=9$, it is also created for large enough $E$ in the toroidal background. Thus, speculation about the Hagedorn transition is not denied in the toroidal background case.

The application of our model to the inflationary cosmology would be very interesting \cite{inflation1} \cite{inflation2}. If $\Eb$ is larger than (\ref{eq:EDd9}) and satisfies the condition (\ref{eq:regionA}) in the early universe then the brane-antibrane pair is stable and the universe expands inflationary. But when $\Eb$ becomes smaller than (\ref{eq:EDd9}) as the universe expands, the brane-antibrane pair becomes unstable and the tachyon starts to roll down from $T=0$ \cite{roll}. It must be noted that if we consider the dilaton gravity, the tension depends on time, because the tension of a D-brane depends on the dilaton field.

Finally, the results that a brane-antibrane pair is created only when it is extended in all the non-compact directions is favorable to the Plank solid model of Schwarzschild black holes \cite{Hotta3}. It might be interesting to study the relation between our model and the black holes.

\section*{Acknowledgements}

The author would like to thank colleagues at Kyoto University for useful discussions.

\end{document}